\documentclass{owrart}

\usepackage{graphicx}

\newtheorem{theorem}{Theorem}

\begin{document}


\begin{talk}{Daniel Ueltschi}
{An improved tree-graph bound}
{Ueltschi, Daniel}

\newcommand{\be}{\begin{equation}}
\newcommand{\ee}{\end{equation}}
\newcommand{\dd}{{\rm d}}
\newcommand{\e}[1]{\,{\rm e}^{#1}\,}
\newcommand{\ii}{{\rm i}}
\newcommand{\Tr}{{\operatorname{Tr\,}}}
\newcommand{\tr}{{\operatorname{tr\,}}}
\newcommand{\Imm}{{\rm Im}\;}
\newcommand{\Ree}{{\rm Re}\;}
\newcommand{\sumtwo}[2]{\sum_{\substack{#1 \\ #2}}}
\newcommand{\prodtwo}[2]{\prod_{\substack{#1 \\ #2}}}

\newcommand{\caC}{{\mathcal C}}
\newcommand{\caT}{{\mathcal T}}
\newcommand{\bbC}{{\mathbb C}}
\newcommand{\bbN}{{\mathbb N}}
\newcommand{\bbR}{{\mathbb R}}
\newcommand{\bbZ}{{\mathbb Z}}
\def\bbone{{\mathchoice {\rm 1\mskip-4mu l} {\rm 1\mskip-4mu l} {\rm 1\mskip-4.5mu l} {\rm 1\mskip-5mu l}}}

\newcommand{\eps}{{\varepsilon}}
\newcommand{\lda}{\langle\!\langle}
\newcommand{\rda}{\rangle\!\rangle}

The day before my talk, Aldo Procacci told us about the beautiful new bound for the convergence of the cluster expansion, which he obtained recently with S.\ Yuhjtman \cite{PY}. This suggested a way to improve the tree-graph bounds stated in \cite{PU}. Additional comments by David Brydges and Tyler Helmuth during Procacci's talk were illuminating; they noticed in particular the relevance of Kruskal's algorithm.

The goals of my talk were to turn the result of \cite{PY} as an explicit tree-graph bound, and to provide a simplified, streamlined proof. It turns out that the result is immediately useful for the work of another participant, Martin Hanke \cite{Han} (who suggested the extension to complex numbers).

The difficulty with the convergence of cluster expansions is to estimate a sum over connected graphs of arbitrary sizes. One needs to use cancellations in order to make it convergent. It turns out that the sum over connected graphs can be reduced to a sum over spanning trees; this sum is considerably smaller.

Here, we state the result with minimal setting. $\caC_n$ and $\caT_n$ denote the sets of connected graphs and of trees with $n$ vertices.

\begin{theorem}
\label{thm}
Let $u_{i,j} \in \bbR$ and $b_i \in [0,\infty)$, $1 \leq i,j \leq n$, be numbers such that for all subsets $I \subset \{1,\dots,n\}$, we have the ``stability condition"
\be
\label{stab}
\sum_{i,j \in I, i<j} u_{i,j} \geq -\sum_{i \in I} b_i.
\ee
Then
\be
\label{bound}
\Bigl| \sum_{g \in \caC_n} \prod_{ij \in g} \bigl( \e{-u_{i,j}} - 1 \bigr) \Bigr| \leq \e{\sum_{i=1}^n b_i} \sum_{t \in \caT_n} \prod_{ij \in t} \bigl( 1 - \e{-|u_{i,j}|} \bigr).
\ee
\end{theorem}

A similar theorem can be found in \cite{PU} with two different upper bounds. The first one follows Ruelle's algebraic method. The second one is motivated by the tree-graph identity of Brydges and Federbush \cite{BF}, combined with an extension of Procacci \cite{Pro}. The two bounds in \cite{PU} are strictly larger than the one above, so this constitutes an improvement indeed.

In the case of complex numbers, $u_{i,j} \in \bbC$, the stability assumption \eqref{stab} is replaced by $\sum \Ree u_{i,j} \geq -\sum_{i\in I} b_i$. One can generalise the tree-graph bound as
\be
\label{complex}
\Bigl| \sum_{g \in \caC_n} \prod_{ij \in g} \bigl( \e{-u_{i,j}} - 1 \bigr) \Bigr| \leq \e{\sum_{i=1}^n b_i} \sum_{t \in \caT_n} \prod_{ij \in t} \bigl| 1 - \e{-|\Ree u_{i,j}| + \ii\, \Imm u_{i,j}} \bigr|.
\ee

Notice that the last term is smaller than $|1 - \e{-u_{i,j}}|$. We now give a proof of Theorem \ref{thm}.

Recall that a {\it partition scheme} is given by a map $T : \caC_n \to \caT_n$ with the property that, for each $t \in \caT_n$, there corresponds a set of edges $E(t)$ such that
\be
T^{-1}(t) = \bigl\{ g \in \caC_n : t \subset g \subset t \cup E(t) \bigr\}.
\ee
(We suppose that $E(t) \cap t = \emptyset$.)

Kruskal's algorithm provides just such a partition scheme. One is given an arbitrary order on all edges of the complete graph of $n$ vertices. Given $g \in \caC_n$, we define a spanning tree by adding edges in increasing order, provided the new edge does not form a loop (if it does, we ignore the new edge). For $t \in \caT_n$, the set $E(t)$ contains exactly all edges $ij \notin t$ such that $ij$ is bigger than all the edges in the path from $i$ to $j$ in $t$. This characterisation of the set $E(t)$ is important.

Given $(u_{i,j})$, we choose an order on edges such that $u_{i,j}$ is nondecreasing. Using Hamlet's lemma ({\it to be or not to be, this is the expansion}), we have
\be
\label{Hamlet}
\begin{split}
& \phantom{\Bigl|} \sum_{g \in \caC_n} \prod_{ij \in g} \bigl( \e{-u_{i,j}} - 1 \bigr) \phantom{\Bigr|} = \sum_{t \in \caT_n} \prod_{ij \in t} \bigl( \e{-u_{i,j}} - 1 \bigr) \prod_{ij \in E(t)} \e{-u_{i,j}}, \\
& \Bigl| \phantom{\sum_{g \in \caC_n} \prod_{ij \in g} \bigl( \e{-u_{i,j}} - 1 \bigr)} \Bigr| \leq \sum_{t \in \caT_n} \prod_{ij \in t} \bigl| \e{-u_{i,j}} - 1 \bigr| \prod_{ij \in E(t)} \e{-u_{i,j}}.
\end{split}
\ee
A key trick in \cite {PY} is to use the identity
\be
\label{trick}
\bigl| \e{-u_{i,j}} - 1 \bigr| = \e{(u_{i,j})_-} \bigl( 1 - \e{-|u_{i,j}|} \bigr).
\ee
The upper bound in Eq.\ \eqref{Hamlet} becomes
\be
\prod_{ij \in t} \bigl| \e{-u_{i,j}} - 1 \bigr| \prod_{ij \in E(t)} \e{-u_{i,j}} = \prod_{ij \in t} \bigl( 1 - \e{-|u_{i,j}|} \bigr) \exp\Bigl\{ -\sum_{ij \in t_-} u_{i,j} - \sum_{ij \in E(t)} u_{i,j} \Bigr\}.
\ee
Here, $t_-$ denotes the set of edges of $t$ where $u_{i,j}<0$. This subgraph is a forest and is illustrated in Fig.\ \ref{fig tree}. 
\begin{figure}[ht]
\includegraphics[width=60mm]{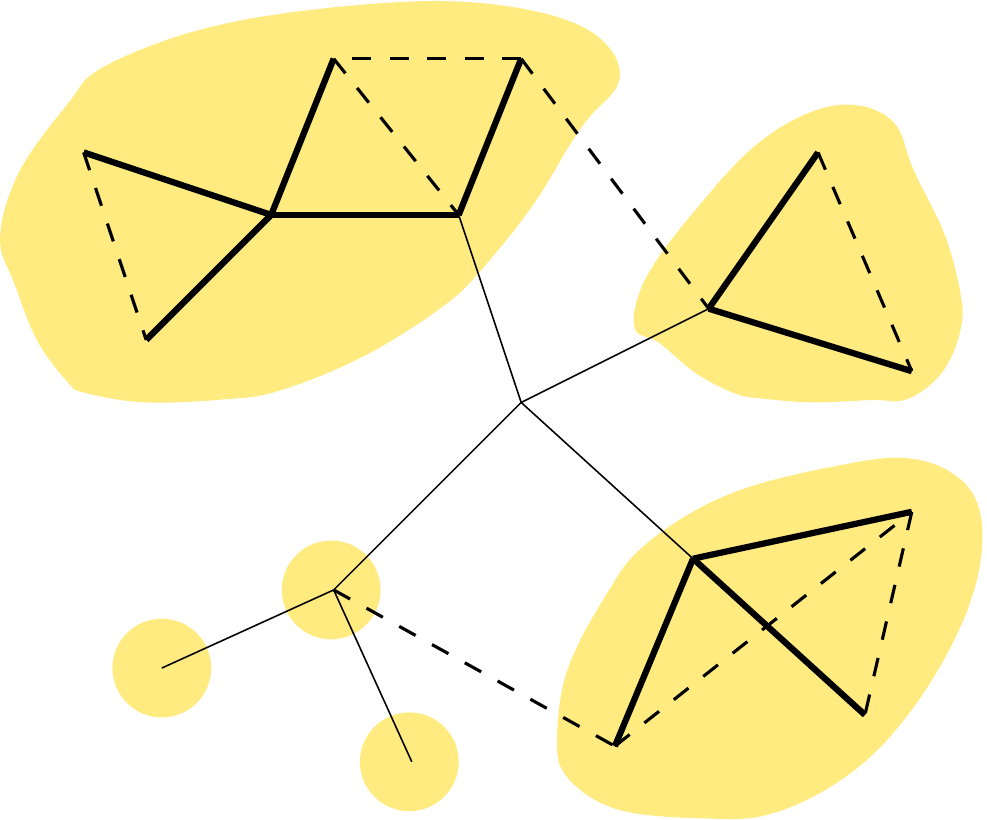}
\caption{The tree $t$ with bold edges when $u_{i,j}<0$ and light edges when $u_{i,j} \geq 0$. Edges of $E(t)$ are shown with dashed lines.}
\label{fig tree}
\end{figure}
Let us denote the forest $\{t_1,\dots,t_k\}$, with $t_m$, $m=1,\dots,k$, being the subtrees. With $K(t_m)$ the complete graph on the vertices of $t_m$, we have the lower bound (which we justify below)
\be
\label{key}
\sum_{ij \in t_-} u_{i,j} + \sum_{ij \in E(t)} u_{i,j} \geq \sum_{m=1}^k \sum_{ij \in K(t_m)} u_{i,j};
\ee
this is larger than $-\sum_{i=1}^n b_i$ by the stability condition. The claim of the theorem follows immediately.

The lower bound \eqref{key} is the clever observation of \cite{PY}. It follows quite easily from the partition scheme of Kruskal's algorithm, because
\begin{itemize}
\item If $ij$ is an edge between distinct subtrees, we necessarily have $u_{i,j} \geq 0$, since it is bigger than at least one nonnegative edge; we neglect them in the lower bound.
\item All positive edges within $K(t_m)$ belong to $E(t)$; indeed, they are bigger than all edges in the path between $i$ and $j$, which are all negative. Thus no extra positive $u_{i,j}$ have been added in the right side.
\item We have perhaps added a few negative $u_{i,j}$ in the right side, which can only make it smaller.
\end{itemize}
This is illustrated in Fig. \ref{fig tree}; this completes the proof of Theorem \ref{thm}.

In the case of complex numbers, we can order the edges according to $\Ree u_{i,j}$; we use $|\e{-u_{i,j}}-1| = \e{-\Ree u_{i,j}} |1 - \e{\Ree u_{i,j} + \ii \, \Imm u_{i,j}}|$ for $ij \in t_-$; then we prove the inequality \eqref{key} with $\Ree u_{i,j}$ instead of $u_{i,j}$, and we get \eqref{complex}.

\end{talk}


\begin{thebibliography}{99}

\bibitem{BF}
D.C. Brydges, P. Federbush,
{\em A new form of the Mayer expansion in classical statistical mechanics},
J. Math. Phys. 19, 2064 (1978)

\bibitem{Han}
M. Hanke,
{\em Fr\'echet differentiability of molecular distribution functions II. The Ursell function},
arXiv:1603.03900 (2016)

\bibitem{PU}
S. Poghosyan, D. Ueltschi,
{\em Abstract cluster expansion with applications to statistical mechanical systems},
J. Math. Phys. 50, 053509 (2009)

\bibitem{Pro}
A. Procacci,
{\em Erratum and Addendum: ``Abstract polymer models with general pair interactions"}, 
J. Stat. Phys. 135, 779 (2009)

\bibitem{PY}
A. Procacci, S. Yuhjtman,
{\em Convergence of Mayer and virial expansions and the Penrose tree-graph identity},
Lett. Math. Phys. 107, 31 (2017)

\end{thebibliography}
\end{document}